\documentclass[12pt,a4paper]{iopart}

\expandafter\let\csname equation*\endcsname=\relax
\expandafter\let\csname endequation*\endcsname=\relax

\usepackage{amsmath,amsfonts,amsthm,amssymb}
\usepackage{color}
\usepackage{graphicx}
\usepackage{hyperref}
\usepackage{subcaption}
\usepackage{tikz}
\usetikzlibrary{arrows}
\usetikzlibrary{calc}
\usetikzlibrary{patterns}
\usetikzlibrary{scopes}

\usepackage{cleveref}

\newcommand{\<}{\langle}
\renewcommand{\>}{\rangle}

\numberwithin{equation}{section}

\begin{document}
  \title{Lattice polymers with two competing collapse interactions}
\author{A Bedini$^1$, A L Owczarek$^1$ and T Prellberg$^2$}
\address{$^1$ Department of Mathematics and Statistics,
  The University of Melbourne, Vic 3010, Australia.}
\address{$^2$ School of Mathematical Sciences, Queen Mary University of London, Mile End Road, London E1 4NS, UK.}
\ead{abedini@ms.unimelb.edu.au, owczarek@unimelb.edu.au, t.prellberg@qmul.ac.uk}

\begin{abstract}
There have been separate studies of the polymer collapse transition, where the collapse was induced by two different types of attraction.
In each case, the configurations of the polymer were given by the same subset of random walks being self-avoiding trails on the square lattice.

Numerical evidence shows that when interacting via nearest-neighbour contacts, this transition is different from the collapse transition in 
square-lattice trails interacting via multiply visited sites. While both transitions are second-order, when interacting via nearest-neighbour 
contacts, the transition is relatively weak with a convergent specific heat, while when interacting via multiply visited sites, the specific heat 
diverges strongly.
Moreover, an estimation of the crossover exponent for the nearest-neighbour contact interaction provides a value close to that of the canonical
polymer collapse model of interacting self-avoiding walks, which also interact via nearest-neighbour contacts.

From computer simulations using the flatPERM algorithm, we extend these studies by considering a model of self-avoiding trails on the square lattice 
containing both types of interaction, and which therefore contains all three of the models discussed above as special cases.
We find that the strong multiply-visited site collapse is a singular point in the phase diagram and corresponds to a higher order multi-critical 
point separating a line of weak second-order transitions from a line of first-order transitions.
\end{abstract}

\maketitle

\section{Introduction}
\label{sec:introduction}

\subsection{The polymer collapse transition}

The collapse transition of a polymer in a dilute solution has been a focus of study in lattice statistical mechanics for decades \cite{DeGennes:1975id,DeGennes:1979uw}. Any lattice model of a collapsing polymer has two key ingredients: an excluded volume effect expressing the impenetrability of monomers, and a short-range attractive force, which mimics the complex monomer-solvent interaction. When the effects of the excluded volume and the short-range attraction balance each other, the polymer undergoes a collapse transition which separates two distinct phases: a swollen and a collapsed phase.

The canonical lattice model of the configurations of a polymer in solution has been the model of self-avoiding walks (SAW) where a random walk on a lattice is not allowed to visit a lattice site more than once. SAW display the desired excluded volume effect and are swollen in size relative to unrestricted random walks at the same length. A common way to introduce a short-range interaction is to assign a negative energy to each non-consecutive pair of monomers lying on neighbouring lattice sites, modelling an effective attractive force. This is the interacting self-avoiding walk (ISAW) model, which is the standard lattice model of polymer collapse using self-avoiding walks.

A different model of a collapsing model can be constructed starting from self-avoiding trails. A self-avoiding trail (SAT) is a lattice walk configuration where the excluded volume is obtained by preventing the walk from visiting the same bond, rather than the same site, more than once. This is a slightly weaker restriction, and SAW configurations are a proper subset of SAT configurations. The interacting version of self-avoiding trails (ISAT), customarily obtained by giving an energy to multiple visited sites, also presents a collapse transition.
It is known that SAW and SAT share the same statistics in their high-temperature phase \cite{Nienhuis:1982dl,Conway:1999dl}, but theoretical prediction and numerical evidence \cite{Owczarek:1995iw,Owczarek:2007dn} strongly suggests that the collapse transition of the ISAT model is in a different universality class to that of ISAW, although there is not completely understood why this would be the case.
To be more specific, it is predicted that the collapse transition in ISAW is a second-order phase transition with a convergent specific heat in accordance with the theoretical description first provided by Duplantier and Saleur \cite{Duplantier:1987bq} as described below. On the other hand, ISAT shows a strongly divergent specific heat at its collapse point.

A recent study \cite{Bedini:2013eg} considered self-avoiding trails interacting via nearest-neighbour contacts (INNSAT) as a hybrid of the two models.
Evidence from computer simulations showed that the collapse transition in INNSAT is different from the collapse transition in ISAT, but similar to ISAW. 
It was also found that the low-temperature phase of the two trail-collapse models differ substantially. The phase associated with multiply visited
site interactions is fully dense in the thermodynamic limit, while the phase associated with nearest-neighbour contacts seems indistinguishable
from the low-temperature phase of interacting self-avoiding walks.

\subsection{Magnetic Systems}

The properties of lattice polymers are also related to those of magnetic systems near their critical point \cite{DeGennes:1975id}. More precisely, the scaling properties of self-avoiding walks are connected to the critical properties of a spin system with $O(n)$ symmetry in the formal limit of zero components ($n \to 0$). From this point of view the collapse transition corresponds to a tri-critical point of such systems \cite{DeGennes:1975id,Duplantier:1982ie}.

Various authors \cite{Nienhuis:1984dj,Nienhuis:1987wc,Jacobsen:2009ut} have studied critical and tri-critical $O(n)$ spin systems. For a special choice of the model on the honeycomb lattice, exact results were obtained in \cite{Nienhuis:1982dl} for two cases: a critical point and a special point governing the low-temperature phase. When $n \to 0$ these two cases become the \textit{dilute} and \textit{dense} polymer phase. 
The dilute and dense phases were also found along two branches of a square-lattice $O(n)$ model \cite{Nienhuis:1984dj,Blote:1989gq} together with two different branches describing the critical behaviour that occurs when $O(n)$ and Ising degrees of freedom on the square lattice display a joint critical point. 

On the other hand, Duplantier and Saleur in 1987 \cite{Duplantier:1987bq} realised that the bond interaction of ISAW could be obtained by introducing vacancies on the honeycomb lattice. Using this observation they could obtain the critical exponents for the polymer collapse transition in the ISAW model. We will refer to the universality class of this critical point as the `$\theta$-point'.
An exact description has now been proposed \cite{Guo:2006fp,Nienhuis:2008ir} for the tri-critical $O(n)$ model in two dimensions as a function of $n$.

When it comes to ISAT the scenario is much less clear, in particular it not obvious how the change of topology caused by the presence of crossings affects the above picture. The description in terms of height model and Coulomb Gas allows  one to consider the presence of crossings only as a perturbation. The exponent associated to loop crossings is the same as that of cubic symmetry breaking, which is known to be irrelevant in the critical $O(n)$ phase, but it has been observed \cite{Jacobsen:2003ib} that this is not true in the low-temperature phase, where the introduction of crossings is a relevant operator which leads to a different universality class. This is generically referred to as the \textit{Goldstone phase} and it is believed \cite{Jacobsen:2003ib,Nienhuis:2008ir} to be described by the intersecting loop model proposed in \cite{Martins:1998di,Martins:1998gp} and since called the Brauer model \cite{Gier:2005ie}.

The relevance of crossings at the tri-critical point (the $\theta$-point when $n \to 0$) is not clear. While the cubic perturbation is still believed to be relevant \cite{Nienhuis:2008ir}, a recent numerical study \cite{Bedini:2013eg} seems to indicate that the $\theta$-point is stable in the presence of crossings, at least with respect to the cross-over and length-scale exponents.

Very recently Nahum {\it et al.\ }\cite{Nahum:2013ei} published a study of loop models with crossings. Their analysis is based on a replica limit of the $\sigma$ model on real projective space $\mathbb{RP}^{n-1}$. They give a field theoretic description of the ISAT which explains the phase diagram found numerically in \cite{Foster:2009cy} and suggests that the ISAT collapse transition is an infinite-order multi-critical point.

\subsection{A model with competing interactions}

In this paper we consider a polymer model of self-avoiding trails with both multiply-visited site and nearest-neighbour interactions. This model generalises ISAT and INNSAT. It also contains in a limiting case the ISAW model, when the Boltzmann weight associated with multiply visited sites is sent to zero.

We study the model via computer simulations using the flatPERM algorithm, and so extend the study of INNSAT in \cite{Bedini:2013eg}.
We point out that this model has been studied some time ago by Wu and Bradley \cite{Wu:1990dw} via real-space renormalisation, which predicted a tetra-critical point 
separating the ISAT and ISAW collapse points.
In contrast, we find that there is likely to be the ISAT collapsed point itself, that separates a line of first-order transitions from ISAW-like weaker $\theta$-point type transitions.

In Section 2 we define more precisely the ISAW, ISAT, and INNSAT models, while in Section 3 we define the model introduced by Wu and Bradley. In Section 4, we present the results of
our simulational studies and deduce a conjectured phase diagram. We end by summarising our conclusions in Section 5.

\section{ISAW and ISAT}
\subsection{Interacting Self-Avoiding Walks (ISAW) and Tri-critical scaling}
\label{sec:isaw}

Let us recall briefly the definition and main properties of the ISAW model.  Consider the ensemble $\mathcal S_n$ of self-avoiding walks (SAW) of length $n$, that is, of all lattice paths of $n$ steps that can be formed on the square lattice such that they never visit the same site more than once.
Given a SAW $\varphi_n \in \mathcal S_n$, we define a \emph{contact} whenever there is a pair of sites that are neighbours on the lattice but not consecutive on the walk. We associate an energy $-\varepsilon_c$ with each contact.
Denoting by $m_c(\varphi_n)$ the number of contacts in $\varphi_n$, the probability of $\varphi_n$ is given by
\begin{equation}
	\frac{e^{\beta \varepsilon_c m_c(\varphi_n)}}{Z_n(T)},
\end{equation}
and the partition function  $Z_n(T)$ is defined in the usual way as
\begin{equation}
	Z_n(T) = \sum_{\varphi_n\in\mathcal S_n}\ e^{\beta \varepsilon_c m_c(\varphi_n)},
\end{equation}
where $\beta$ is the inverse temperature $1/k_B T$ ($k_B$ is Boltzmann's constant).
We define a Boltzmann weight (fugacity) $\omega_c = \exp(\beta
\varepsilon_c)$.  The finite-length reduced free energy is
\begin{equation}
  \kappa_n(T) = \frac{1}{n} \log\ Z_n(T)
  \label{eq:free-energy-1}
\end{equation}
and the thermodynamic limit is obtained by taking the limit of large $n$, i.e.,
\begin{equation}
  \kappa(T) = \lim_{n \to \infty} \kappa_n(T).
  \label{eq:free-energy-2}
\end{equation}
As mentioned above, it is expected that there is a collapse phase transition at a temperature $T_\theta = 1.4986(11)$ \cite{Caracciolo:2011iz}, which is known as the $\theta$-point, characterised by a non-analyticity in $\kappa(T)$. 

The temperature $T_\theta$ also separates regions of different finite-length scaling behaviour for fixed temperatures. Considering this finite-length scaling, for high temperatures ($T > T_\theta$) the excluded volume interaction is the dominant effect, and the behaviour
is universally the same as for the non-interacting SAW problem: for large $n$, the mean squared end-to-end distance (or equivalently the radius of gyration) $R_n^2$ and partition function $Z_n$ are expected to scale as
\begin{align}
  R_n^2 & \sim A \,n^{2 \nu} \quad \text{with $\nu > 1/2$}\quad\text{ and} \\
  Z_n   & \sim D\, \mu^n n^{\gamma-1},
\end{align}
respectively, where the exponents $\nu$ and $\gamma$ are expected to be universal. The constants $A$ and $D$ are temperature dependent and the \emph{connective constant} $\mu$ is related to the free energy; indeed from \cref{eq:free-energy-1,eq:free-energy-2} we have $\log \mu = \kappa(T)$.

The two-dimensional case has been extensively studied by Coulomb-gas techniques and conformal field theory (CFT). It is well established that the universal exponents are those of the dilute polymer universality class \cite{Nienhuis:1982dl}, $\nu=3/4$ and $\gamma=43/32$, for $T>T_\theta$, and those predicted by Duplantier and Saleur \cite{Duplantier:1987bq}, $\nu = 4/7$ and $\gamma = 8/7$, for $T = T_\theta$

For low temperatures ($T < T_\theta$) it is accepted that the partition function is dominated by globular configurations that are internally dense, though not necessarily fully dense. The partition function should then scale differently from that at high temperatures and one expects a large-$n$ asymptotics of the form
\begin{align}
	R_n^2 & \sim A \,n\quad\text{and} \\
	Z_n   & \sim D \,\mu^{n} \mu_s^{\sqrt n} n^{\gamma-1},
\end{align}
where the constants $A$ and $D$ are again temperature dependent. The factor $\mu_s^{\sqrt n}$ (with $\mu_s < 1$ ) takes into account the surface contribution to the free energy \cite{Owczarek:1993bk}. Indeed $\sqrt{n}$ is the average number of steps on the boundary of the globule (with the boundary having fractal dimension equal to 1). It is expected that the internal density smoothly goes to zero as the temperature is raised to the $\theta$-point. Collapsed polymers in two dimensions are expected to fall in the universality class of dense polymers \cite{Nienhuis:1982dl} although numerical simulations have excluded the predicted value 97/46 for exponent $\gamma$ \cite{Duplantier:1993kz,Baiesi:2006bt}.

To explore the singularity in the free energy at the collapse point further, it is useful to consider the (reduced) internal energy and the specific heat, which are defined as
\begin{equation}
	u_n(T) = \frac{\< m_c \>}{n} \quad \mbox{ and } \quad
	c_n(T) = \frac{\<m_c^2\> - \<m_c\>^2 }{n}
	,
\end{equation}
with limits
\begin{align}
	U(T) = \lim_{n \to \infty} u_n(T) \quad \mbox{ and } \quad
	C(T) = \lim_{n \to \infty} c_n(T)
	.
\end{align}
When $T\to T_\theta$, the singular part of the specific heat behaves as
\begin{equation}
	C(T) \sim B\, \left|T_\theta - T\right|^{-\alpha} ,
\end{equation}
where $\alpha < 1$ for a second-order phase transition. If the
transition is second-order, the singular part of the thermodynamic
limit internal energy behaves as
\begin{equation}
	U(T) \sim B\, \left|T_\theta - T\right|^{1-\alpha}
\end{equation}
when $T\to T_\theta$, and there is a jump in the internal energy at
$T_\theta$ if the transition is first-order (an effective value of
$\alpha = 1$).

Tri-critical scaling \cite{Lawrie:1984ta,Brak:1993ci} predicts that around the critical temperature, the finite-length scaling of the singular part of the specific-heat $c_n(T)$ obeys the following crossover scaling form
\begin{equation}
	c_n(T) \sim n^{\alpha \phi}\ \mathcal C\big( (T-T_\theta) n^{\phi} \big) ,
\end{equation}
when $T\to T_\theta$ and $n\to\infty$, and that the exponents $\alpha$
and $\phi$ are related via
\begin{equation}
	2 - \alpha = \frac{1}{\phi}.
\end{equation}
If one considers the peak of the finite-length specific heat it will
behave as
\begin{equation}
	c_n^{peak} \sim n^{\alpha \phi}\ \mathcal C\big( x^{max}),
\end{equation}
where $x^{max}$ is the location of the maximum of the function
$\mathcal{C}(x)$.

The predicted exponents for the $\theta$-point collapse \cite{Duplantier:1987bq} are
\begin{equation}
  \phi = \phi_{\theta} = 3/7 \quad \text{and} \quad \alpha = \alpha_{\theta}= - 1/3 .
\end{equation}
It is important to observe that this implies that the specific heat
does not diverge at the transition since the exponent
\begin{equation}
  \alpha_{\theta} \phi_{\theta} =-1/7\approx - 0.14
\end{equation}
is negative. However, the peak values of the third derivative of the
free energy with respect to temperature will diverge with positive
exponent
\begin{equation}
  (1 + \alpha_{\theta}) \phi_{\theta} = 2/7\approx0.28.
\end{equation}

\subsection{Interacting Self-Avoiding Trails (ISAT)}
\label{sec:isat}

The model of interacting trails on the square lattice is defined as
follows. Consider the ensemble $\mathcal T_n$ of self-avoiding trails
(SAT) of length $n$, that is, of all lattice paths of $n$ steps that
can be formed on the square lattice such that they never visit the
same bond more than once. Note that $\mathcal S_n \subset \mathcal T_n$.
Given a SAT $\psi_n \in \mathcal T_n$, we associate an energy
$-\varepsilon_t$ with each doubly visited site.  Denoting by
$m_t(\psi_n)$ the number of doubly visited sites in $\psi_n$, the
probability of $\psi_n$ is given by
\begin{equation}
	\frac{e^{\beta \varepsilon_t m_t(\psi_n)}}{Z^{\text{\sc ISAT}}_n(T)},
\end{equation}
where we define the Boltzmann weight $\tau = \exp(\beta
\varepsilon_t)$ and the partition function of the ISAT model is given
by
\begin{equation}
	Z^{\text{\sc ISAT}}_n(T) = \sum_{\psi_n\in \mathcal T_n}\ \tau^{m_t(\psi_n)}.
\end{equation}
Previous work \cite{Owczarek:1995iw,Owczarek:2007dn} on the square
lattice has shown that there is a collapse transition at a temperature
$T=T_t$ with a strongly divergent specific heat, and the exponents
have been estimated as
\begin{equation}
  \phi_{\text{\sc it}} =0.84(3)\quad \mbox{ and } \quad \alpha_{\text{\sc it}}=0.81(3)\;,
\end{equation}
arising from a scaling of the peak value of the specific heat
diverging with exponent
\begin{equation}
    \alpha_{\text{\sc it}} \phi_{\text{\sc it}}=0.68(5)\;.
\end{equation}
This result is a clear difference to the ISAW $\theta$-point described
above where the singularity in the specific heat is \emph{convergent}.
Additionally, at $T=T_t$ the finite-length scaling of the end-to-end
distance was found to be consistent \cite{Owczarek:1995iw} with the
form
\begin{equation}
  R_n^2(T) \sim A n\left(\ln n\right)^2
\end{equation}
as $n\to\infty$. Again, this is quite different to the exponent
$\nu=4/7$ for the ISAW.

Another important difference that has been recently observed \cite{Doukas:2010bb,Bedini:2013kd} is that the low temperature phase is maximally dense. On the square lattice this implies that if one considers the proportion of the sites on the trail that are at lattice sites which are not doubly occupied via
\begin{equation}
  p_n=\frac{n - 2 \< m_t \>}{n},
  \label{eq:p-n-def}
\end{equation}
then it is expected that
\begin{equation}
  p_n \rightarrow 0 \quad \mbox{as} \quad n\rightarrow \infty.
\end{equation}
Very recently the ISAT model has been studied in the context of a loop model with crossings \cite{Nahum:2013ei}, where it has been suggested that the ISAT $\theta$-point is an infinite-order multi-critical point described by the $O(n \to 1)$ sigma model studied in \cite{Jacobsen:2003ib}. 

\subsection{Nearest-Neighbour Interacting Self-Avoiding Walks (INNSAT)}
\label{sec:nnisat}

The Nearest-Neighbour Interacting Self-Avoiding Walks (INNSAT) is a third model,  recently studied in \cite{Bedini:2013eg}, specifically crafted to mix features of the ISAW and ISAT models. The model is defined as follows: consider the set of bond-avoiding paths $\mathcal T_n$ as defined in the previous section. When two sites are adjacent on the lattice but not consecutive along the walk, so as not to be joined by {\it any} step of the walk, we again refer to this pair of sites as a nearest-neighbour \emph{contact} and we give it a weight $\omega_c=e^{\beta \varepsilon_c}$, analogously to the ISAW model. Denoting by $m_c(\psi_n)$ the number of contacts in $\psi_n$, the probability of $\psi_n$ is given by
\begin{equation}
	\frac{e^{\beta \varepsilon_c m_c(\psi_n)}}{Z^{\text{\sc nt}}_n(T)},
\end{equation}
where the partition function is
\begin{equation}
  Z^{\text{\sc nt}}_n(T) = \sum_{\psi_n\in\mathcal T_n}\ \omega_c^{m_c(\psi_n)}.
\end{equation}
The intensive reduced internal energy  and specific heat are as for ISAW:
\begin{equation}
	u_n(T) = \frac{\< m_c \>}{n} \quad \mbox{ and } \quad
	c_n(T) = \frac{\<m_c^2\> - \<m_c\>^2 }{n} .
\end{equation}
We shall also consider the proportion of the sites on the trail that
are at lattice sites which are not doubly occupied via the quantity $p_n$ defined in \eqref{eq:p-n-def}.

One would think that the presence of crossings would affect the universality class of the collapse transition (e.g. as portrayed in \cite{Nahum:2013ei}) but in \cite{Bedini:2013eg} it was shown that the INNSAT model has a collapse transition in the same universality class as ISAW, that is the $\theta$-point.

\begin{figure}[b]
  \centering
  \includegraphics{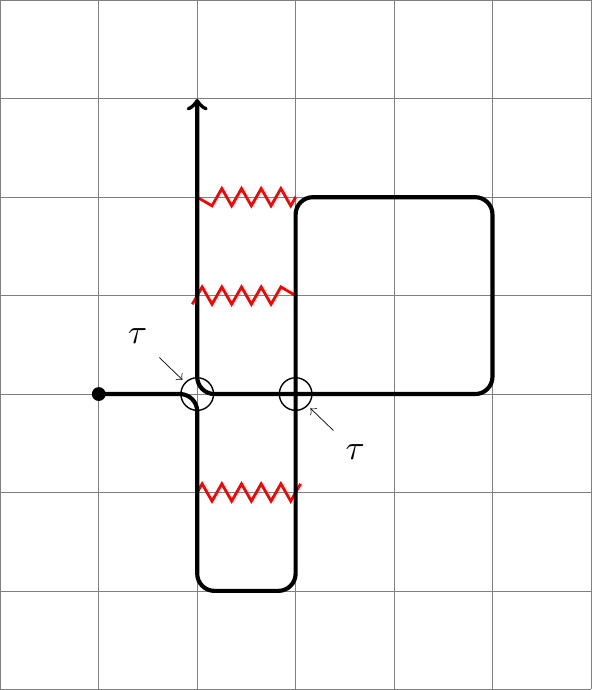}
  \caption{An example of a configuration for the two interaction model, with $m_c = 3$ (as there are $3$ nearest-neighbour contacts illustrated via zigzag (red) lines) and $m_t = 2$ (as there are 2 doubly-visited sites). The trail can visit a site of the lattice twice by ``touching" and by ``crossing" itself. Note that there is no contact between the second and the seventh visited site of the walk, even though these are non-consecutive nearest-neighbour sites, as both sites are visited consecutively by a different segment of the trail. }
\label{fig:configuration}
\end{figure}

\section{The Wu-Bradley model.}
\label{sec:isat-nn}

A model of ISAT with nearest-neighbour interaction can be defined as follows. Consider the set of bond-avoiding paths $\mathcal T_n$ as defined in the previous section. Given a SAT $\psi_n \in \mathcal T_n$, we associate an energy $-\varepsilon_t$ every time the path visits the same site more than once, as in ISAT. Additionally, we define a \emph{contact} whenever there is a pair of sites that are neighbours on the lattice but not consecutive on the walk. We associate an energy $-\varepsilon_c$ with each contact.

For each configuration $\psi_n \in \mathcal T_n$ we count the number $m_t(\psi_n)$ of doubly-visited sites and $m_c(\psi_n)$ of contacts: see Figure~\ref{fig:configuration}. Hence we associate with each configuration a Boltzmann weight $\tau^{m_t(\psi_n)} \omega^{m_c(\psi_n)}$ where $\tau = \exp(\beta \varepsilon_t)$, $\omega = \exp(\beta \varepsilon_c)$, and $\beta$ is the inverse temperature $1/k_B T$. The
partition function of the model is given by
\begin{equation}
  Z_n(\tau, \omega) = \sum_{\psi_n\in\mathcal T_n}\
  \tau^{m_t(\psi_n)} \omega^{m_c(\psi_n)}
  .
\end{equation}
The probability of a configuration $\psi_n$ is then
\begin{equation}
  p(\psi_n; \tau, \omega) = \frac{ \tau^{m_t(\psi_n)}
    \omega^{m_c(\psi_n)} }{ Z_n(\tau, \omega) }
  .
\end{equation}
The average of any quantity $Q$ over the ensemble set of paths $\mathcal T_n$ is given generically by
\begin{equation}
  \langle Q \rangle(n; \tau, \omega) = \sum_{\psi_n\in\mathcal
    T_n} Q(\psi_n) \, p(\psi_n; \tau, \omega)
  .
\end{equation}
In particular, we can define the average number of doubly-visited
sites per site and their respective fluctuations as
\begin{equation}
  u^{(t)} = \frac{ \langle m_t \rangle }{n} \quad \mbox{ and } \quad
  c^{(t)} = \frac{ \langle m_t^2 \rangle - \langle m_t \rangle^2 }{n}
  .
\end{equation}
One can also consider the average number of contacts of the trail and their fluctuations
\begin{equation}
  u^{(c)} = \frac{ \langle m_c \rangle }{n} \quad \mbox{ and } \quad
  c^{(c)} = \frac{ \langle m_c^2 \rangle - \langle m_c \rangle^2 }{n}
  .
\end{equation}
This model interpolates between three previously-studied models; when we set $\omega = 1$ the model reduces to the ISAT model. If otherwise we set $\tau = 0$ doubly visited sites are excluded and the model reduces to the ISAW model. Finally if we set $\tau = 1$ it becomes the INNSAT model studied in \cite{Bedini:2013eg}.

\section{Numerical results}

We began by simulating the full two parameter space by using the flatPERM algorithm \cite{prellberg2004a-a}. FlatPERM outputs an estimate $W_{n,\mathbf{k}}$ of the total weight of the walks of length $n$ at fixed values of some vector of quantities $\mathbf{k}=(k_1,k_2,\dotsc,k_{\ell})$. From the total weight one can access physical quantities over a broad range of temperatures through a simple weighted average, e.g.
\begin{align}
  \< \mathcal O \>_n(\tau) = \frac{\sum_{\mathbf{k}} \mathcal O_{n,\mathbf{k}}\,
   \left(\prod_j \tau_j^{k_j}\right) \, W_{n,\mathbf{k}}}{\sum_\mathbf{k} \left(\prod_j \tau_j^{k_j}\right) \, W_{n,\mathbf{k}}}.
\end{align}
The quantities $k_j$ may be any subset of the physical parameters of
the model. To study the full two parameter phase space we set $(k_1, k_2) = (m_t, m_c)$ and $(\tau_1, \tau_2) = (\tau, \omega)$.

\begin{figure}[ht!] \centering
  \includegraphics[width=0.5\linewidth]{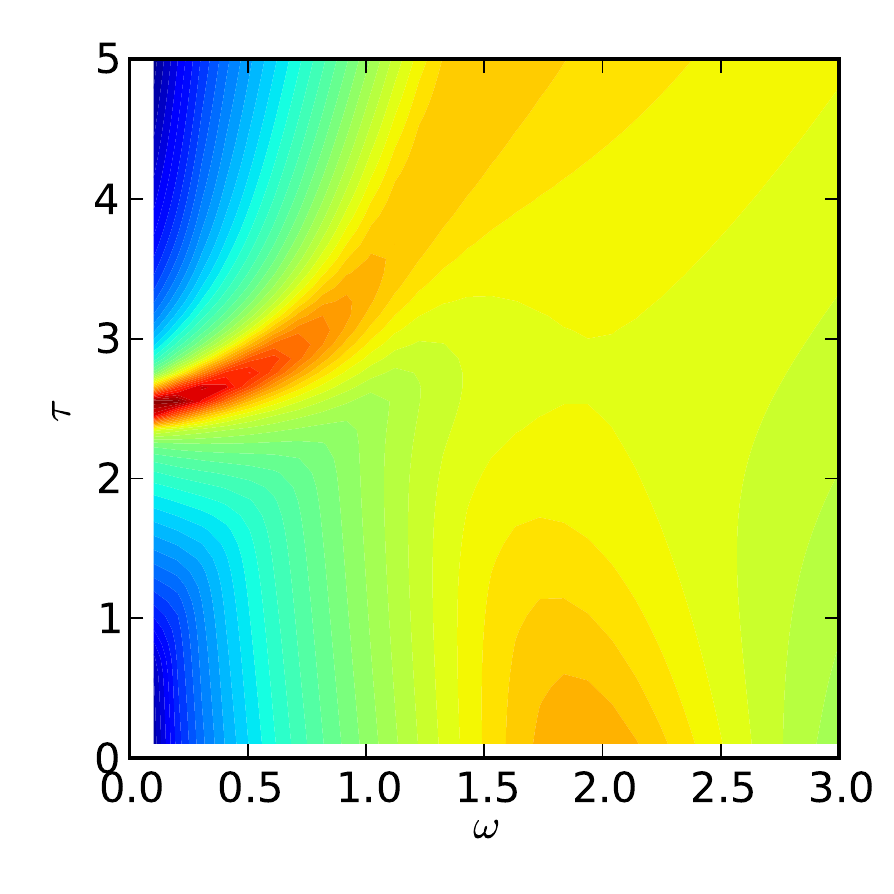}
  \includegraphics[width=0.5\linewidth]{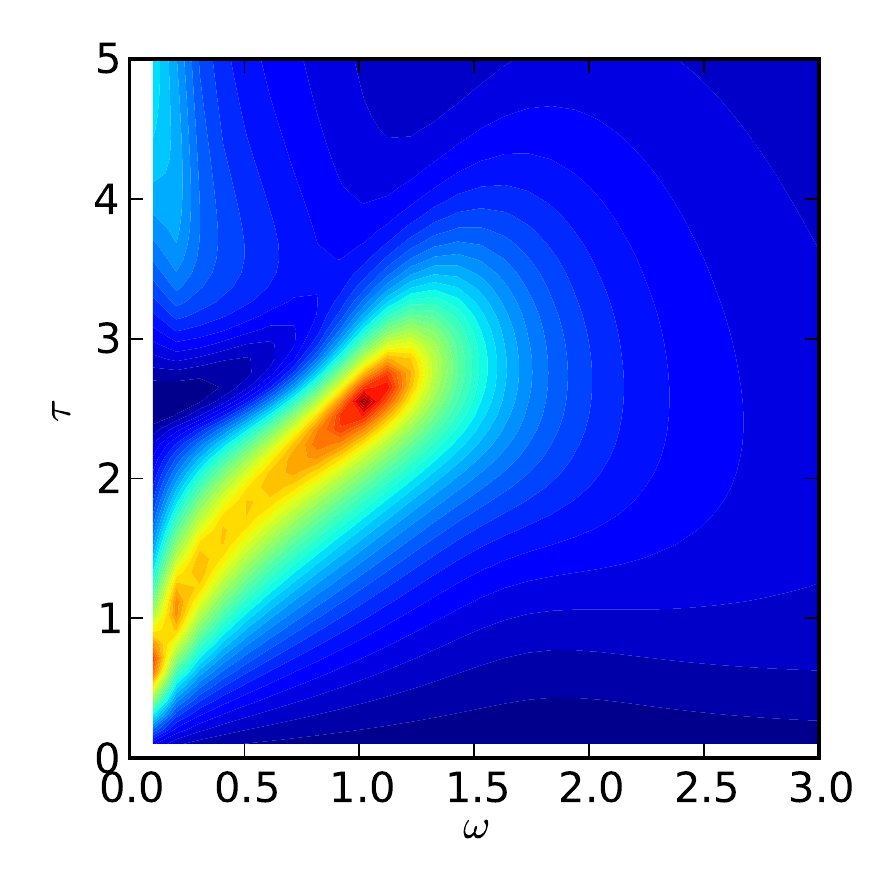}

  \caption{Left: density plot of the logarithm of the largest eigenvalue $\lambda_{max}$ of the matrix of second derivatives of the free energy with respect to $\tau$ and $\omega$ at length 256. Right: density plot of the ratio $\lambda_{min}/\lambda_{max}$ of the eigenvalues of the matrix of second derivatives of the free energy with respect to $\tau$ and $\omega$ at length 256.}
  \label{fig:eigenvalue_plot}
\end{figure}

We have first simulated the model using the full two-parameter flatPERM algorithm up to length $n = 256$, running $4.4 \cdot 10^6$ iterations, and collecting $2.3 \cdot 10^{11}$ samples at the maximum length.
To obtain a landscape of possible phase transitions, we plot the largest eigenvalue of the matrix of second derivatives of the free energy with respect to $\tau$ and $\omega$ (measuring the strength of the fluctuations and covariance in $m_t$ and $m_c$) at length $n=256$ on the left-hand side of Figure~\ref{fig:eigenvalue_plot}.

\begin{figure}[ht!] \centering
  \includegraphics[width=0.9\linewidth]{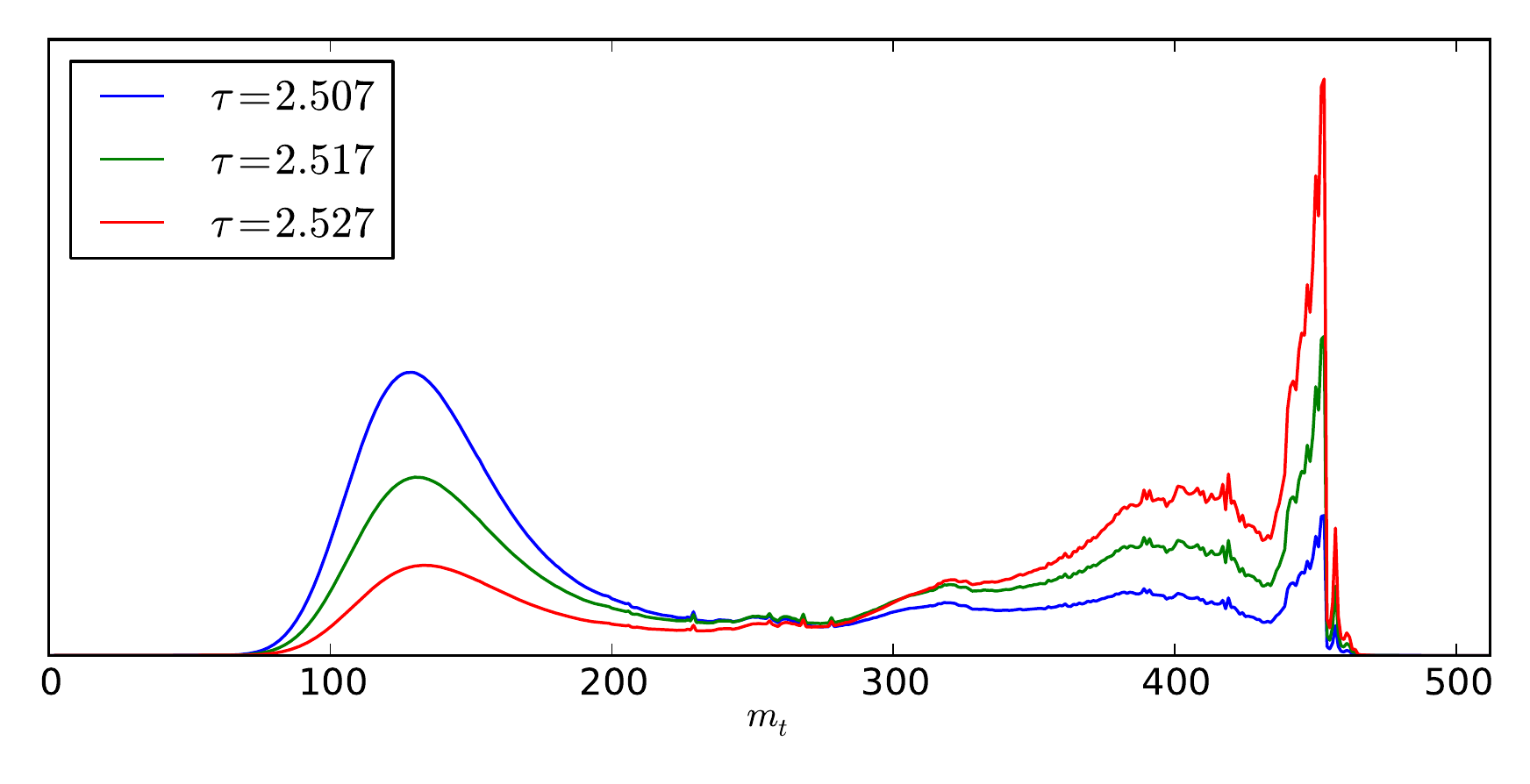}
\caption{Distribution of the number of contacts for various $\tau$ at fixed $\omega = 0.5$.}
\label{fig:double_peaks}
\end{figure}

We expect the ISAT transition to be somewhat shifted away from $(\omega,\tau)=(1,3)$ due to considering a finite-size ensemble. If one considers the vertical line $\omega=1$, one notices that the maximal eigenvalue peaks at a value of $\tau$ somewhat greater than $3$. We remind that the phase transition for $\omega=1$ is a strong second-order phase transition where the specific heat diverges with an exponent $0.68(5)$. For $\omega<1$, there exists a line of even stronger peaks that join with the peak at $\omega=1$, from which one can infer that there exists a strong phase transition on varying $\tau$ for each $\omega<1$. This is borne out by finite-size scaling analysis. Specifically, we have studied the model when $\omega=0.5$:
for this value of $\omega$ we have simulated the model using a one-parameter flatPERM algorithm up to length $n = 1024$, running $7.9 \cdot 10^6$ iterations, and collecting $2.7 \cdot 10^{10}$ samples at the maximum length. We find that the specific heat divergence is commensurate with a first-order transition with a linear divergence. To test this assumption of a first-order transition, we consider the distribution of the number of contacts for various values of $\tau$ near the peak of the specific heat. Figure \ref{fig:double_peaks} shows a clear bimodal distribution, confirming the first-order character of the transition.

It is hence likely that there exists a line of first-order phase transitions at values of $\tau$ near $3$ for each value of $\omega\lesssim 1$. We will return to the question of whether the line of first-order transitions extends all the way to $\omega=1$ below.

Returning to our two-parameter data, let us first note that when $\tau=0$ our model is the ISAW model and hence there is a weak $\theta$-like transition at $(\omega,\tau)\sim(1.94,0)$ \cite{Caracciolo:2011iz}, which is reflected in the density plot by a broad peak near $\omega = 2$ on varying $\omega$ when $\tau=0$. A line of such weak peaks extends to larger values of $\tau$. When $\tau=1$, the model becomes the previously studied INNSAT \cite{Bedini:2013eg}, which also demonstrated a weak $\theta$-like transition with exponent estimates encompassing ISAW values. We thus conjecture that the entire line lies in the $\theta$-universality class. 

Now if the suggestion that the ISAT collapse corresponds to an infinite order multi-critical point \cite{Nahum:2013ei} is correct, it is then natural and simplest to conjecture that the line of first-order transitions meets with the line of $\theta$-like transitions at that point.

We obtain an indication of where the two lines might join by considering the ratio between the two eigenvalues of the covariance matrix. This is based on the idea that the two transitions are driven by the two different types of interaction. When the two eigenvalues coincide is then argued to indicate the presence of a higher order critical point.  The density plot of the eigenvalue ratio is shown on the right-hand side of Figure~\ref{fig:eigenvalue_plot}. One clearly observes a unique point close to the ISAT collapse point, where the two eigenvalues have the same magnitude.

\begin{figure}[ht!] \centering
  \includegraphics[width=0.9\linewidth]{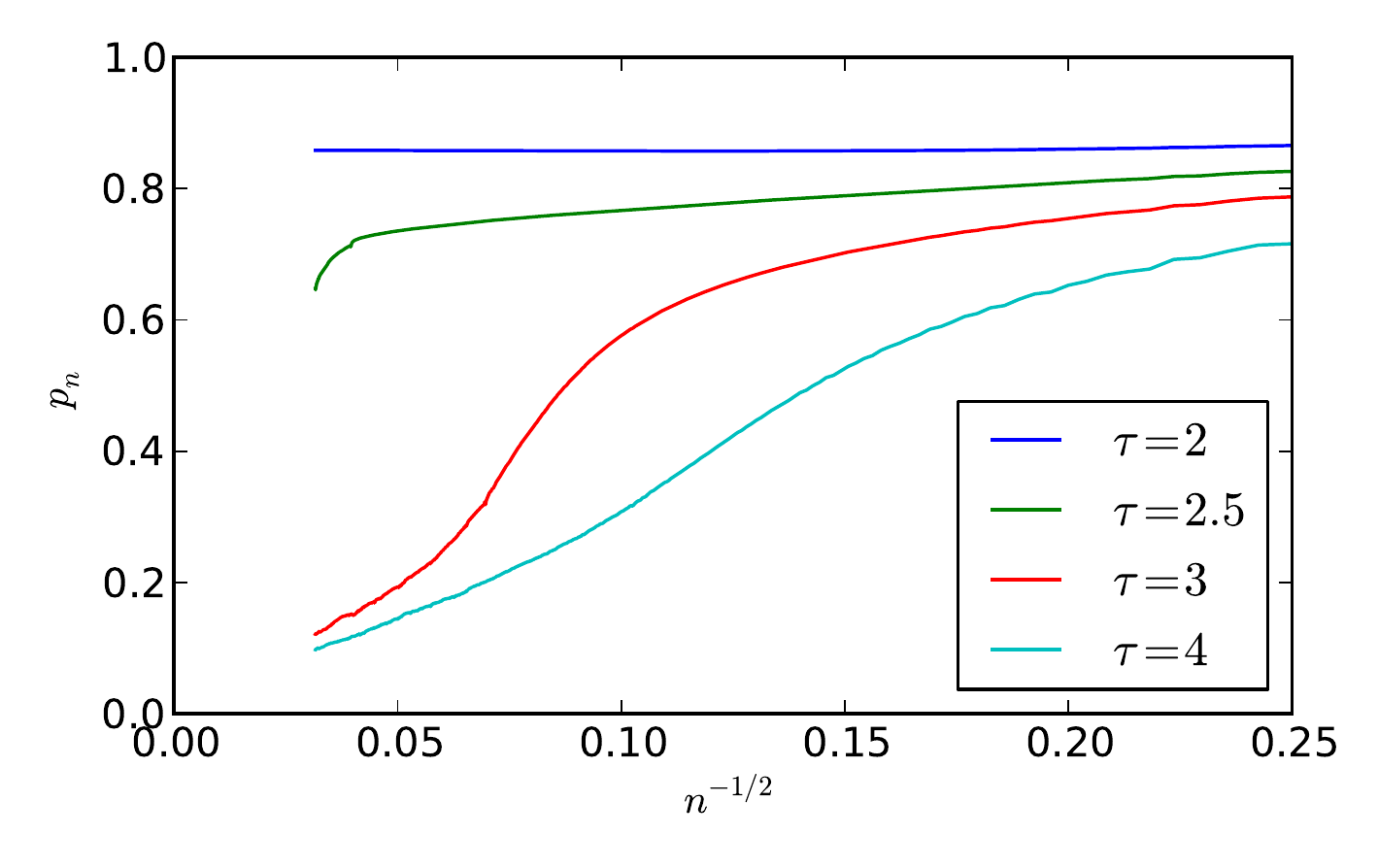}
  \caption{Plots of $p_n$, the proportion of steps that visit singly-occupied sites at fixed $\omega=0.5$ and $\tau=2.0$, $2.5$, $3.0$, and $4.0$ (from top to bottom), versus $n^{-1/2}$. The scale $n^{1/2}$ is the natural scale for the border of a dense configuration.}
  \label{fig:low-temperature}
\end{figure}

Considering again the density plot of the largest eigenvalue, the line of peaks that is associated with first-order transitions for $\omega <1$ and that meets the ISAT critical point for $\omega =1$ extends to higher values of $\omega$. This implies some type of phase transition at low temperatures. To understand what this transition might be, we now consider the low temperature phases for fixed values of $\omega$ and $\tau$. When $\omega=1$, an analysis of the ISAT model has previously shown that the low temperature phase is \emph{maximally dense} with the proportion $p_n$ of sites on the trail that are not doubly occupied going to zero in the thermodynamic limit of infinite length \cite{Doukas:2010bb,Bedini:2013kd}. Here we plot the same quantity for various temperatures when $\omega =0.5$ in Figure~\ref{fig:low-temperature}.
At low values of $\tau$ the quantity $p_n$ converges to a non-zero value while for larger values $p_n$ seems to converge to zero within error, with a transition visible around $\tau\approx2.5$. 

\begin{figure}[ht!] \centering
  \subcaptionbox*{$m_t = 132$}
  {\includegraphics[width=0.4\linewidth]{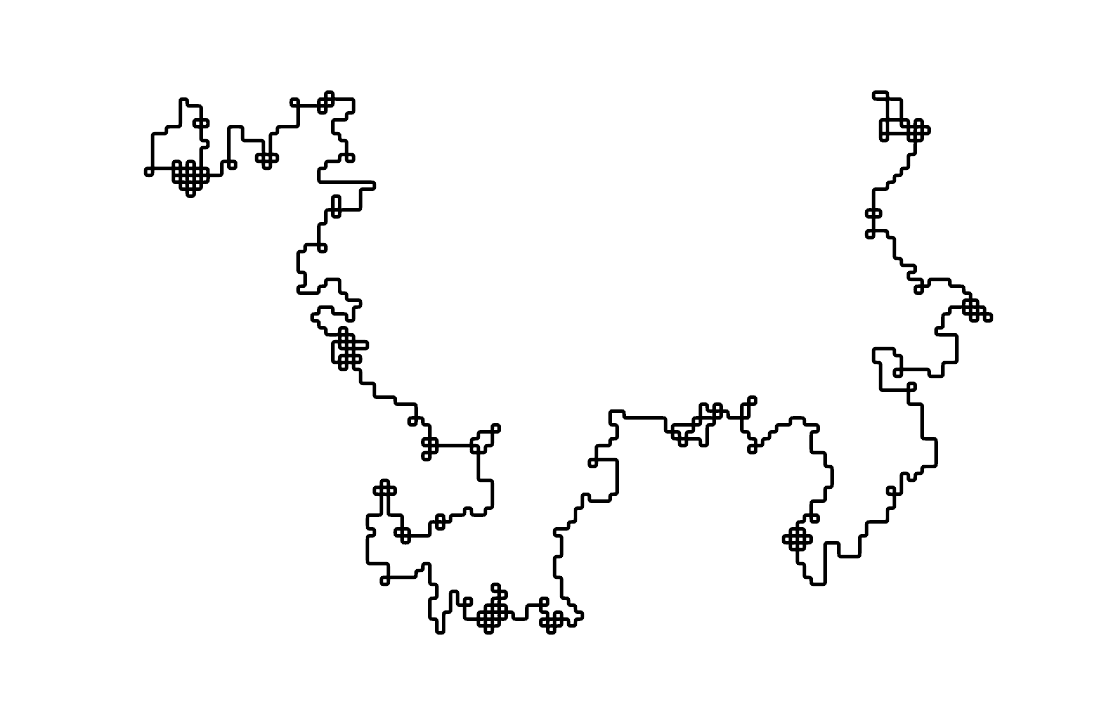}}
  \subcaptionbox*{$m_t = 256$}
  {\includegraphics[width=0.4\linewidth]{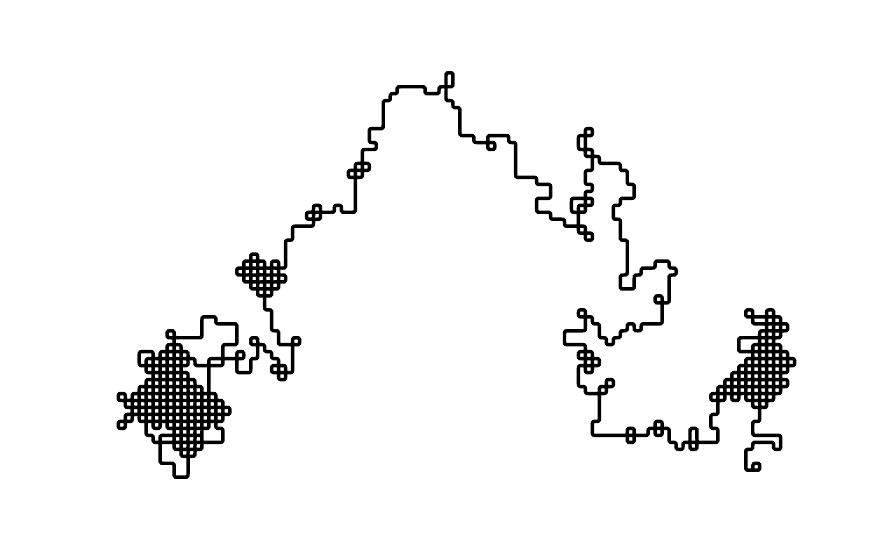}}
  \subcaptionbox*{$m_t = 350$}
  {\includegraphics[width=0.4\linewidth]{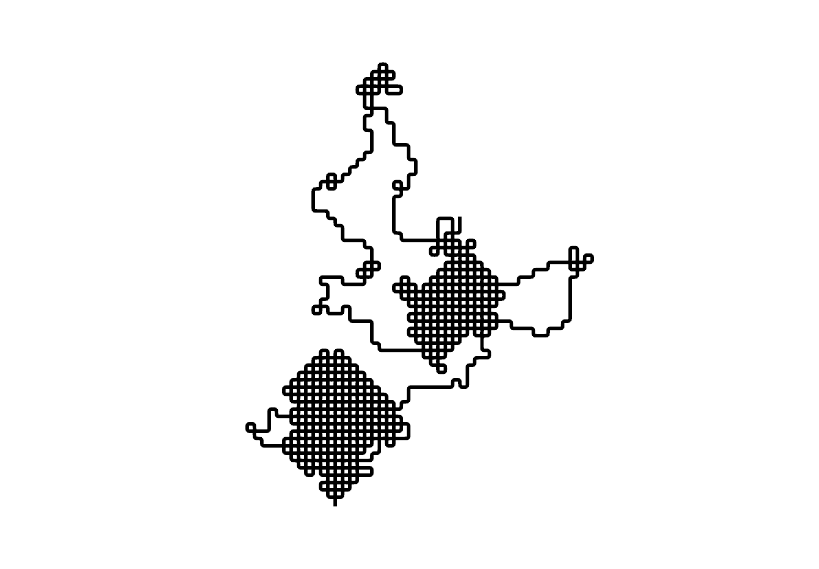}}
  \subcaptionbox*{$m_t = 400$}
  {\includegraphics[width=0.4\linewidth]{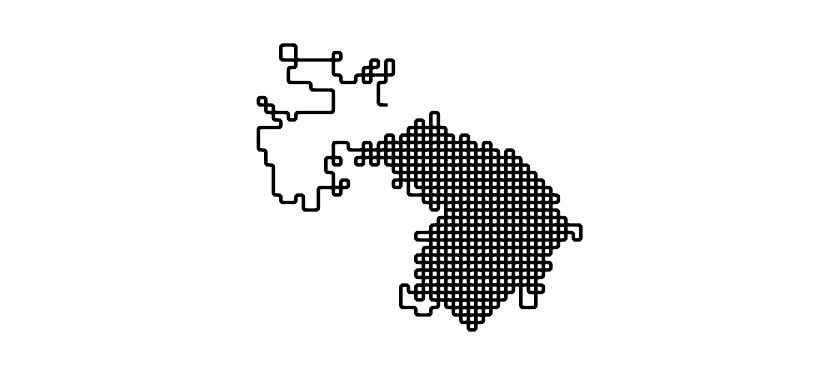}}
  \subcaptionbox*{$m_t = 452$}
  {\includegraphics[width=0.4\linewidth]{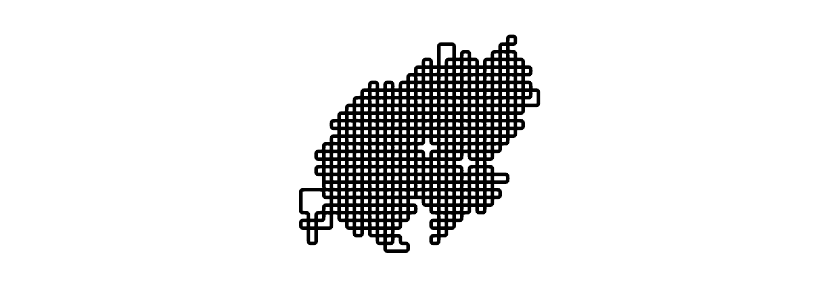}}
\caption{Typical configurations of $1024$-step walks with specified numbers $m_t$ of contacts, that have been generated in a simulation at $\omega=0.5$. The values of $m_t$ have been chosen such as to cover the whole range of the histogram in Figure \ref{fig:double_peaks}. The configurations illustrate the co-existence of fully dense and swollen parts of the polymer, demonstrating the first-order nature of the transition.}
\label{fig:configs}
\end{figure}

To illustrate the nature of the polymer around the transition, we present some typical configurations with specified numbers of contacts that have been generated in a simulation at $\omega=0.5$. These can be seen in Figure~\ref{fig:configs}.
Not only do these configurations illustrate the nature of the low-temperature phase where the number of contacts is large (see the configuration with $m_t=452$), but they also clearly demonstrate the first-order nature of the collapse, as we observe co-existence of fully dense and swollen parts of the polymer for smaller values of $m_t$.

Now let us consider low temperatures for fixed values of $\tau$. In our previous work on INNSAT \cite{Bedini:2013eg}, when $\tau=1$ the quantity $p_n$ was seen to converge to a non-zero value regardless of temperature: see Figure~7 in \cite{Bedini:2013eg}. This phase is unambiguously of a different nature as that for large $\tau$ at fixed $\omega$. We can therefore conclude that the line of peaks for large values of $\omega$ and $\tau$ in the largest eigenvalue plot is associated with a transition between these two low temperature phases: one being maximally dense and the other not. We have not investigated this transition here but an analysis of a transition between similar phases \cite{Krawczyk:2009xyz} leads us to conjecture that it is second order.

\begin{figure}[ht!] \centering
  \includegraphics[width=0.5\linewidth]{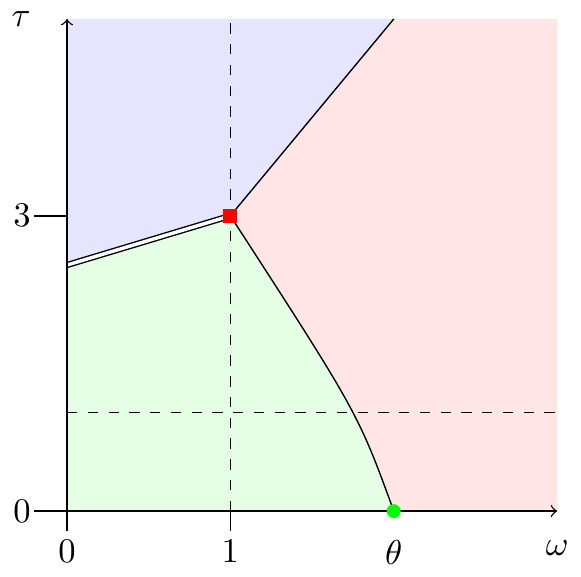}

  \caption{Proposed phase diagram. The double-line indicates a line of first-order transitions. The horizontal and vertical dashed lines are respectively the ISAT and INNSAT model, the ISAW model correspond to the axis $\tau = 0$. The red square and the green circle correspond respectively to the ISAT collapse point and the ISAW $\theta$-point.}
  \label{fig:phase-diagram}
\end{figure}

The entire phase diagram in the $\omega,\tau$-quadrant is therefore split into three phases we already know: the swollen high-temperature phase in the lower-left corner, a low-temperature globular phase of the ISAW model on the right (as determined in \cite{Bedini:2013eg}), and the low-temperature maximally dense phase of the ISAT model in the top left. We then have two critical lines and one first order line joining together at the multi-critical point located at $(\omega,\tau)=(1,3)$ that separates these phases. Putting all this information together gives us the conjectured phase diagram in Figure~\ref{fig:phase-diagram}.

\section{Conclusions}

We investigated a two-parameter model of polymer collapse that has competing interactions, first studied by Wu and Bradley \cite{Wu:1990dw}, and contains three previously investigated models of polymer collapse, namely ISAT, ISAW and INNSAT, as specialisations.

We find that the phase diagram  for the Wu-Bradley model contains three phases: a swollen phase and two collapsed phases, one of which is maximally dense.
The corresponding three phase boundaries, two of which are second order and one of which is first-order, meet seemingly at the collapse point of the ISAT model, and is in agreement with the suggestion in \cite{Nahum:2013ei} that the ISAT transition is an infinite-order multi-critical point.
The second-order phase transition line between the swollen and not maximally dense collapsed phase contains the transitions in the ISAW and INNSAT models, which we conjecture to be a line of $\theta$-like transitions.

The existence of the maximally dense phase is believed to be related to the $O(n \to 0)$ Goldstone phase \cite{Jacobsen:2003ib,Nahum:2013ei}. 
It is an open question why there exist two different low-temperature regions, as the presence of crossings is supposed to lead to the same Goldstone phase \cite{Nahum:2013ei}.
More work needs to be done to elucidate the nature of these phases, and to resolve the transition between them. At present, this seems to be out of reach of available algorithms.

\section*{Acknowledgements}

Financial support from the Australian Research Council via its support
for the Centre of Excellence for Mathematics and Statistics of Complex
Systems and through its Discovery Program is gratefully acknowledged
by the authors. A L Owczarek thanks the School of Mathematical
Sciences, Queen Mary, University of London for hospitality. We thank
Adam Nahum for helpful discussions.
  
\section*{References}

\providecommand{\newblock}{}

\end{document}